\documentclass[prl,article,nofootinbib,twocolumn,preprintnumbers,superscriptaddress,amsmath,amssymb,aps,longbibliography]{revtex4-1}

\usepackage{graphicx}
\usepackage{dcolumn}
\usepackage{bm}
\usepackage[table]{xcolor}
\usepackage{multirow}
\usepackage{comment}
\usepackage{url}
\usepackage{epsfig}
\usepackage{amsfonts}
\usepackage{amsmath}
\usepackage{graphicx}
\usepackage{color}
\usepackage{amssymb,amsmath,hyperref,slashed,color,graphicx,bm,cancel,url}
\usepackage[normalem]{ulem}
\usepackage[T1]{fontenc}
\usepackage{relsize}
\hypersetup{colorlinks,citecolor= niceblue,linkcolor= niceblue, urlcolor=niceblue}
\definecolor{niceblue}{rgb}{0.1,0.2,0.6}
\definecolor{kjkblue}{rgb}{0.39, 0.589, 0.6914}

\DeclareMathAlphabet{\mathpzc}{OT1}{pzc}{m}{it}

\graphicspath{{Figures/}}

\begin{document}

\preprint{FERMILAB-PUB-20-576-T, N3AS-20-005, NUHEP-TH/20-12}

\title{Intimate Relationship Between Sterile Neutrino Dark Matter and $\Delta N_{\rm eff}$}

\author{Kevin J. Kelly}
\email{kkelly12@fnal.gov, ORCID: 0000-0002-4892-2093}
\affiliation{Theory Department, Fermi National Accelerator Laboratory, Batavia, IL 60510, USA}
\author{Manibrata Sen}
\email{manibrata@berkeley.edu, ORCID: 0000-0001-7948-4332
}
\affiliation{Department of Physics, University of California Berkeley, Berkeley, CA 94720, USA}
\affiliation{Department of Physics \& Astronomy, Northwestern University, Evanston, IL 60208, USA}
\author{Yue Zhang}
\email{yzhang@physics.carleton.ca, ORCID: 0000-0002-1984-7450
}
\affiliation{Department of Physics, Carleton University, Ottawa, ON K1S 5B6, Canada}

\date{\today}

\begin{abstract}
    The self-interacting neutrino hypothesis is well motivated for addressing the tension between the origin of sterile neutrino dark matter and indirect detection constraints. It can also result in a number of testable signals from the laboratories to the cosmos. We show that, in a broad class of models where the sterile neutrino dark matter relic density is generated by a light neutrinophilic mediator, there must be a lower bound on the amount of extra radiation in early universe, in particular $\Delta N_{\rm eff}>0.12$ at the CMB epoch. This lower bound will be further strengthened with an improved $X$-ray search at the Athena observatory. Such an intimate relationship will be unambiguously tested by the upcoming CMB-S4 project.
\end{abstract}

\pacs{Valid PACS appear here}
\maketitle

\textbf{Introduction --}
The origin of dark matter (DM) poses a fascinating puzzle. A sterile neutrino with a keV-scale mass and small mixing with active neutrinos is a simple, longstanding DM candidate. Along with other traditional candidates (e.g., WIMPs/axions), sterile neutrinos are highly testable. In the minimal setup, the active-sterile neutrino mixing that can account for the observed DM relic density~\cite{Dodelson:1993je} is already in tension with searches for DM decaying into monochromatic $X$-rays~\cite{Abazajian:2017tcc}. An elegant solution for alleviating this tension is to introduce a new self interaction among active neutrinos. This keeps neutrinos in thermal equilibrium with themselves longer in the early Universe and facilitates efficient sterile neutrino DM (S$\nu$DM) production~\cite{deGouvea:2019phk, Kelly:2020pcy}. This mechanism is attractive because it predicts a number of novel signatures for laboratory tests~\cite{Berryman:2018ogk, deGouvea:2019qaz, Brdar:2020nbj}. 
Interestingly, the self-interacting neutrino scenario emerges from a broad class of well-motivated beyond-Standard-Model frameworks~\cite{Chikashige:1980qk, Gelmini:1980re,Das:2017iuj, Kelly:2019wow, Ng:2014pca, Ioka:2014kca, Forastieri:2015paa, Forastieri:2019cuf,Kreisch:2019yzn, Blinov:2019gcj, Shalgar:2019rqe, Bustamante:2020mep, Berryman:2014yoa, Coloma:2017zpg, deGouvea:2019goq, Chacko:2019nej, Chacko:2020hmh}. 

We will soon enter a new precision era of cosmology. The Cosmic Microwave Background - Stage 4 (CMB-S4) project~\cite{Abazajian:2016yjj} is expected to measure cosmological parameters to an unprecedented level. In particular, it could restrict the amount of extra radiation (in units of extra neutrinos, $\Delta N_{\rm eff}$) during the CMB epoch to a few-percent uncertainty, while it is still allowed (perhaps even favored) to be order one by the current data~\cite{Ade:2015xua,Bernal:2016gxb}. This could shed light on existing tensions resulting from various ways of measuring the Hubble parameter~\cite{Freedman:2017yms, Wong:2019kwg, Riess:2019cxk, Aghanim:2018eyx}, as well as fundamental theories that can accommodate deviations from the standard cosmology. Moreover, $\Delta N_{\rm eff}$ measurement during the big-bang nucleosynthesis (BBN) epoch puts further restrictions on room for new physics.

In this letter, we point out an intriguing relationship between the two important ingredients of our Universe -- the origin of S$\nu$DM and $\Delta N_{\rm eff}$ for both CMB and BBN.
The connection is established by a light mediator with a feeble coupling to active neutrinos such that it enters the thermal bath after BBN but decays away before CMB. 
S$\nu$DM then is dominantly produced from active neutrino self scattering through the exchange of on-shell mediator. 
The post-BBN thermalization of the mediator inevitably modifies the standard cosmology predictions by making new contributions to $\Delta N_{\rm eff}^\text{\sc cmb}$ and $\Delta N_{\rm eff}^\text{\sc bbn}$.
As a result, the viable parameter space for S$\nu$DM relic density is highly correlated with the predictions of $\Delta N_{\rm eff}$. 
We further point out a new implication of the upcoming $X$-ray search by the Athena experiment for the next-generation precision measurements in cosmology-- this scenario connects different frontiers of cosmological and astrophysical probes.

\textbf{Models --}
S$\nu$DM is a linear combination of a gauge singlet fermion and the active neutrino states from the Standard Model~\cite{Dodelson:1993je}, $\nu_4 = \nu_s \cos\theta + \nu_a \sin\theta$,
where $\nu_4$ denotes a mass eigenstate and  $\theta$ is the active-sterile mixing angle in vacuum. The sterile neutrino is a decaying DM candidate whose longevity is attributed to its small mass/mixing. We introduce a new force among active neutrinos,
\begin{equation}\label{eq:PhiIntLagrangian}
    \mathcal{L}_{\rm int.} = \frac{\lambda_{\alpha\beta}}{2}\nu_\alpha \nu_\beta \phi + \mathrm{h.c.} \ , \quad (\alpha, \beta = e, \mu, \tau),
\end{equation}
which can arise from dimension-six gauge invariant operators at energy scales well below the weak scale. 
The scalar field $\phi$ could be either real or complex. If real, $\phi$ could be identified as the Majoron~\cite{Chikashige:1980qk, Gelmini:1980re}, where $\lambda = i M_\nu/f$ and $f$ is the spontaneous lepton number breaking scale. If complex, $\phi$ also contains the radial mode and restores $B-L$ symmetry at high scales~\cite{Berryman:2018ogk}. 
The impact of active neutrino self-interactions on the relic density of S$\nu$DM was first explored in~\cite{deGouvea:2019phk}.
The new force enables more frequent active neutrino scattering than normal weak interactions, thereby, S$\nu$DM can be produced with a smaller mixing angle than required by Dodelson-Widrow~\cite{Dodelson:1993je}. Ref.~\cite{deGouvea:2019phk} focuses on $m_\phi$ between MeV--GeV, where the new interaction is sufficiently strong to fully thermalize $\phi$ before BBN.

\begin{figure}[t]
  \includegraphics[width=0.95\linewidth]{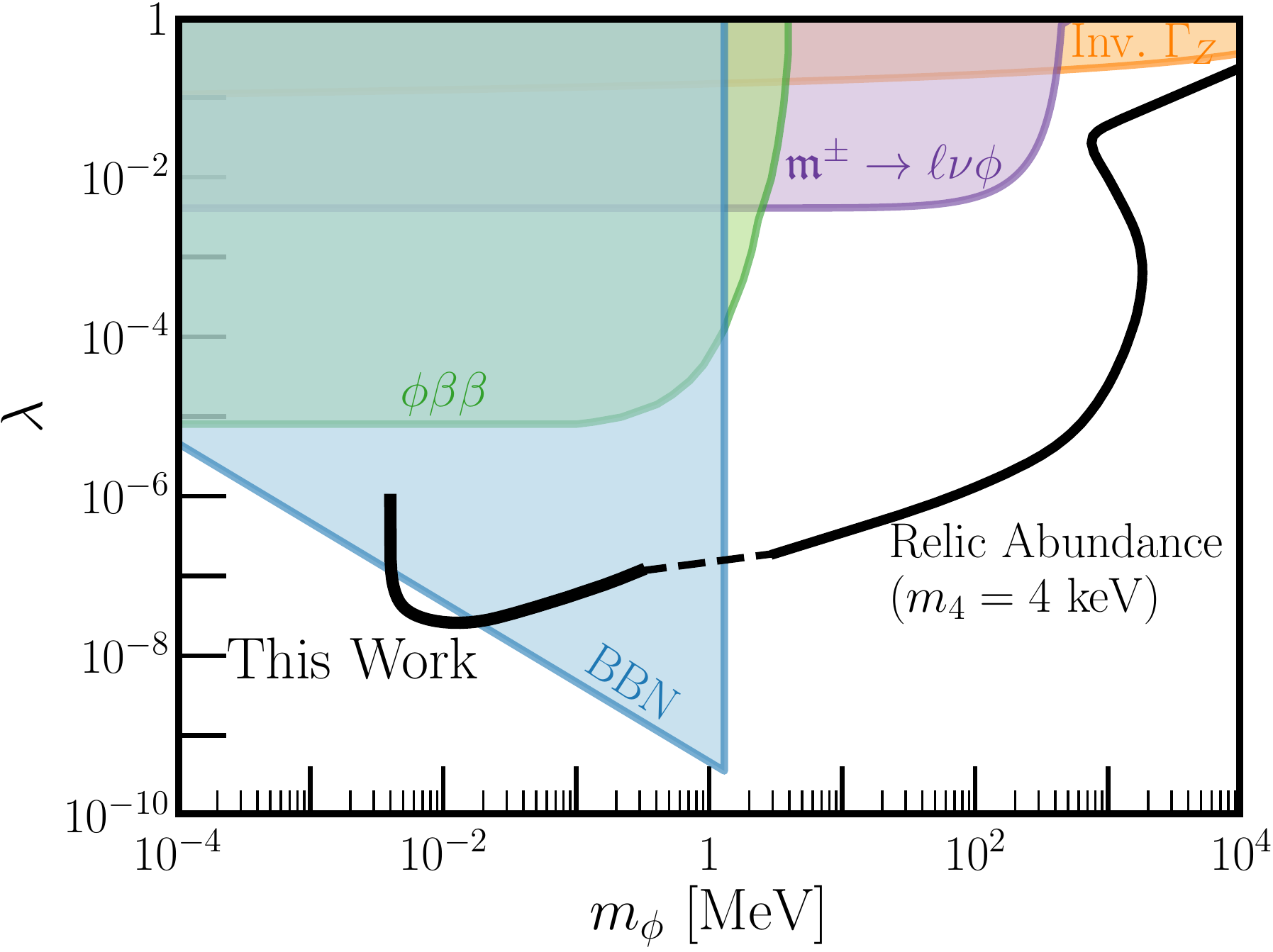}
    \caption{The Big Picture: full parameter space that can account for the relic density of S$\nu$DM (thick black curve). We choose $m_4=4$\,keV and $\sin^22\theta = 5\times 10^{-9}$.
    The colored regions show the experimental constraints on neutrino self-interaction mediated by a real scalar $\phi$, including leptonic meson decays~\cite{Pasquini:2015fjv, Berryman:2018ogk, Blinov:2019gcj, deGouvea:2019qaz} (purple), invisible $Z$ width~\cite{Berryman:2018ogk, Brdar:2020nbj} (orange), double beta decay~\cite{Agostini:2015nwa,Blum:2018ljv,Brune:2018sab} (green), and $\Delta N_{\rm eff}$ for successful BBN~\cite{Pitrou:2018cgg} (blue). We will explore the S$\nu$DM production mechanism in the the sub-MeV $m_\phi$ region marked by ``This Work'' which could be fully tested by the upcoming CMB-S4 and Athena experiments. 
    }
    \label{fig:LambdaMPhi:Broad}
\end{figure}

The present work broadens the focus and opens up the possibility of a sub-MeV mediator $\phi$. The constraint from BBN is lifted if $\phi$ does not thermalize before/during the BBN epoch and has no primordial population from other high-scale physics~\cite{Shi:1998km, Shaposhnikov:2006xi, Asaka:2006ek, Roland:2014vba, Hansen:2017rxr, Johns:2019cwc, Bezrukov:2009th, Nemevsek:2012cd, Dror:2020jzy}. This is achieved with a feeble coupling to neutrinos and a phase space suppression for $\nu\nu\to\phi$ to occur at temperatures much higher than $m_\phi$. The condition for not thermalizing $\phi$ before BBN sets an upper bound on $\lambda$ that scales as $1/m_\phi$, whereas correct DM relic density requires $\lambda$ to scale as $\sqrt{m_\phi}$ in the small coupling regime. As a result, viable parameter space opens up for $m_\phi$ between keV and MeV scales.

Remarkably, our result, in tandem with that from Ref.~\cite{deGouvea:2019phk}, nicely forms the ``Big Picture'' of Fig.~\ref{fig:LambdaMPhi:Broad}.
The thick black curve depicts the entire range of the theory space where the relic density of S$\nu$DM is explained, aided by neutrino self-interaction.
Viable mediator mass range divides into two windows, above and below MeV scale. They correspond to different histories of the neutrino and S$\nu$DM sector in the early universe.
In the heavier mediator region, the strongest probes are from terrestrial experiments.
In contrast, the sub-MeV mediator regime will be best explored by cosmological observations of $\Delta N_{\rm eff}$, to be detailed below.

\textbf{Cosmology --}
S$\nu$DM production is governed by
\begin{equation}\label{eq:Boltzmann}
    \frac{d f_{4}(E, z)}{dz} = \frac{\Gamma(E, z) \sin^22\theta}{4 H z} f_{a}(E, z) \Theta(E - m_{4}) \ ,
\end{equation}
where and $f_4$/$f_a$ are phase space distribution functions of sterile/active neutrinos, respectively. We introduce $z \equiv m_\phi/T_\nu^\text{\sc sc}$ to label time, where $T_\nu^\text{\sc sc}$ is defined as the neutrino temperature in the standard cosmology without $\phi$; $z$ expands linearly with the Universe's radius. The actual neutrino temperature $T_\nu$ is affected by the interaction in Eq.~(\ref{eq:PhiIntLagrangian}). After weak-interaction decoupling, $T_\nu$ deviates from $T_\nu^\text{\sc sc}$ as the $\phi$ population builds. $\Gamma$ is the thermally-averaged scattering rate for an active neutrino, and $H$ is the Hubble parameter. Since we are interested in small $m_\phi$ and $\lambda$, the effective active-sterile mixing angle is approximately equal to the vacuum one throughout the dominant S$\nu$DM production window.

The distribution function $f_{a}$, after active neutrinos thermalize with $\phi$, is characterized by $T_\nu$ and a chemical potential $\mu_\nu$, 
\begin{equation}\label{eq:activeNudistr}
f_{a}(E, z) = \frac{1}{1 + \exp\left[\left(E - \mu_\nu(z)\right)/T_\nu(z) \right]} \ .
\end{equation}
To derive their $z$ dependence, we introduce three time scales,
\begin{itemize}
\item $z_A$: the onset of BBN -- neutrinos have just decoupled from weak interaction. We assume $\phi$ has not yet equilibrated with neutrinos;
\item $z_B$: the $\nu$-$\phi$ system has just equilibrated. In general, $\phi$ remains ultra-relativistic;
\item $z_C$: all $\phi$ have decayed away ($T\ll m_\phi$), mostly into neutrinos, with a small fraction decaying into S$\nu$DM.
\end{itemize}

Before $z_A$, neutrinos and photons obey $T_\nu = T_\nu^\text{\sc sc} =T_\gamma$. The $\nu$-$\phi$ interaction freezes in a sub-thermal population of $\phi$ via the Boltzmann equation 
\begin{equation}\label{eq:PhiFreezeIn}
\frac{d Y_\phi}{dz} = - \frac{\gamma_{\phi \leftrightarrow \nu \nu}}{s H z} \left( \frac{Y_\phi}{Y_\phi^\text{\sc eq}}  - 1 \right) \ ,
\end{equation}
where $Y_\phi=n_\phi/s$ is the yield, $n_\phi$ is the number density of $\phi$, and $s$ is the entropy density of the Universe. The thermally-averaged rate is defined as $\gamma_{\phi \rightarrow \nu \nu} 
\equiv n_\nu^{eq} \langle (m_\phi/E) \Gamma_{\phi\rightarrow \nu \nu} \rangle$ where $\Gamma_\phi = 3\lambda^2 m_\phi/(32\pi)$ is the decay rate at rest. 
Here we assume $\phi$ couples universally to all neutrinos, $\lambda_{ee,\mu\mu,\tau\tau}=\lambda$. 
 In the small $m_\phi$ limit,
$\gamma_{\phi \rightarrow \nu \nu}  \simeq m_\phi^3 \Gamma_\phi /(2\pi^2 z^2)$.\footnote{The $\nu\bar\nu\to\phi\phi^{(*)}$ process is only important for the coupling $\lambda$ much larger than those of interest to this work~\cite{Huang:2017egl, Barenboim:2020dmg}.}
Assuming no $\phi$ population at early times, integrating Eq.~\eqref{eq:PhiFreezeIn} up to $z_A$ determines the population of $\phi$ and its contribution to $\Delta N_{\rm eff}^\text{\sc bbn}$.

\begin{figure*}
  \includegraphics[width=0.7\linewidth]{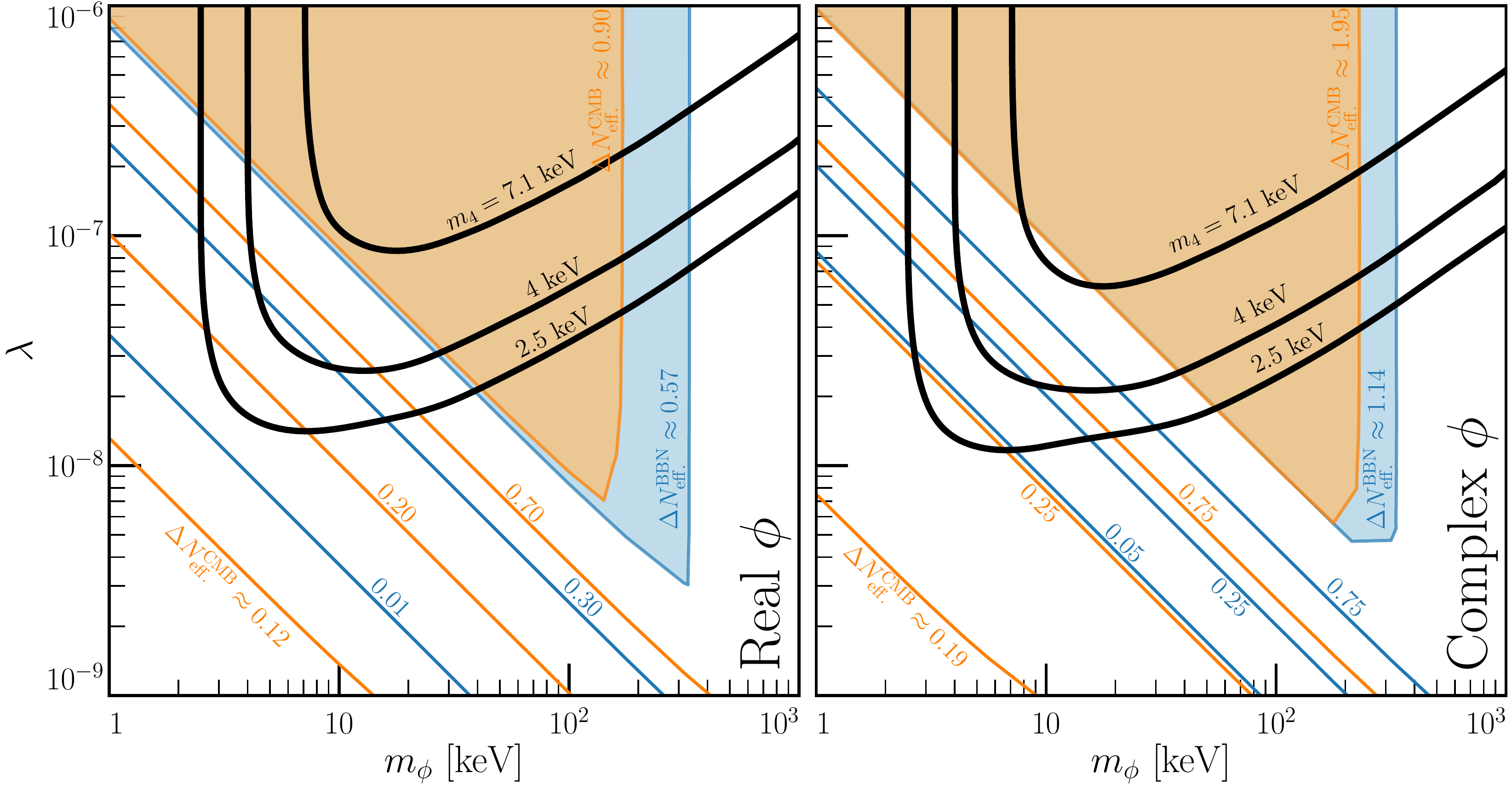}
    \caption{Sterile neutrino DM relic density (black contours) and $\Delta N_{\rm eff}$ (orange and blue contours for CMB and BBN, respectively), as a function of $\lambda$ and $m_\phi$ for real (left) and complex (right) scalar $\phi$. We present three choices of the S$\nu$DM mass with the largest experimentally allowed mixing angles (see main text). 
}
    \label{fig:LambdaMPhi}
\end{figure*}

From $z_A$--$z_B$, the $\nu$-$\phi$ interaction induces chemical equilibrium. At $z_B$, $T_\nu$ deviates from $T_\nu^\text{\sc sc}$; a chemical potential is also generated, $\mu_\phi =2  \mu_\nu=2 \mu_{\bar\nu}$ for real scalar $\phi$, or $\mu_\phi =2 \mu_\nu$ and $\mu_{\phi^*}=2 \mu_{\bar\nu}$ for complex $\phi$. Without CP violation, $\mu_\phi=\mu_\phi^*$. At $z_B$, $T_\nu$ and $\mu_\nu$ are dictated by energy/lepton-number conservation~\cite{Chacko:2003dt, Escudero:2019gvw,Escudero:2020dfa, Berlin:2017ftj}. Entropy is not preserved here; the $\phi$ thermalization process is irreversible. In our supplemental material, we provide details of this matching.  

The most important epoch for S$\nu$DM production is from $z_B$--$z_C$, by the $\nu\nu\to\nu \nu_4$ process with on-shell, $s$-channel $\phi$ exchange. This is terminated after $\phi$ becomes non-relativistic and freezes out. Entropy/lepton-number conservation allow us to derive $T_\nu(z)$ and $\mu_\nu(z)$. The total energy of the Universe increases because some $\phi$ decay while turning non-relativistic. 
The neutrino self-interaction rate also depends on $T_\nu(z)$ and $\mu_\nu(z)$, 
\begin{equation}
\begin{split}
\Gamma(E, z) &= \frac{3\lambda^2 m_\phi^2 T_\nu}{8 \pi E^2} 
\left[ \ln \left( e^{\mu_\nu/T_\nu} + e^{\omega} \right) - \omega \rule{0mm}{4mm}\right]\sqrt{1-\frac{m_4^2}{m_\phi^2}} \ ,
\end{split}
\end{equation}
where $\omega = {m_\phi^2}/({4 E T_\nu})$. With the above, we integrate Eq.~\eqref{eq:Boltzmann} up to $z_C$ to obtain the relic abundance of S$\nu$DM,
\begin{equation}\label{eq:relicdensity}
\Omega_{4}= \frac{m_{4} s_0}{2\pi^2 s(z_C) \rho_0} \int_{m_{4}}^\infty E dE \sqrt{E^2 - m_{4}^2} f_{4} (E, z_C) \ ,
\end{equation}
where $s_0=2891.2\,{\rm cm^{-3}}$ and $\rho_0 = 1.05\times10^{-5}\, h^{-2}\,{\rm GeV/cm^{-3}}$ are entropy and critical energy densities today~\cite{Zyla:2020zbs}.
The total entropy density at time $z_C$ (we take $z_C = 30$) is
\begin{equation}
s(z_C) = \frac{2\pi^2}{45} \left[ 2 T_\gamma^3 + 2 s_\nu\left(T_\nu(z_C), \mu_\nu(z_C)\rule{0mm}{4mm}\right) \rule{0mm}{5mm}\right] \ ,
\end{equation}
where $T_\gamma \simeq 1.39 T_\nu^\text{\sc sc} = 1.39 m_\phi/z_C$, and 
\begin{equation}
s_\nu(T_\nu, \mu_\nu) = -\frac{3}{\pi^2} \left[ 4 {\rm Li}_4(-e^{\mu_\nu/T_\nu}) - \frac{\mu_\nu}{T_\nu} {\rm Li}_3(-e^{\mu_\nu/T_\nu}) \right] \ .
\end{equation}

The $\nu$-$\phi$ dynamics also allow one to calculate $\Delta N_{\rm eff}^\text{\sc bbn}$ and $\Delta N_{\rm eff}^\text{\sc cmb}$. A positive $\Delta N_{\rm eff}^\text{\sc bbn}$ is generated by the non-thermal production of $\phi$ at temperatures above the MeV scale, 
\begin{equation}\label{eq:NeffBBN}
\Delta N_{\rm eff}^\text{\sc bbn} \simeq 3.046 \left(\frac{ \rho_\phi(z_A)}{\rho_\nu(z_A)}\right) \ ,
\end{equation}
where $\rho_\nu(z_A) = {9\zeta(3)} T_\nu(z_A)^4/({4\pi^2})$ and $\rho_\phi(z_A)$ can be calculated by solving Eq.~(\ref{eq:PhiFreezeIn}). The thermal reaction rate $\gamma_{\phi \leftrightarrow \nu \nu}$ is proportional to $(\lambda m_\phi)^2$, implying weaker BBN constraints for light $\phi$.

By CMB times, $\phi$ particles have decayed away. The contribution to $\Delta N_{\rm eff}^\text{\sc cmb}$ is due to the non-standard distribution function of active neutrinos,~\footnote{S$\nu$DM production via active-sterile mixing also leads to a slight reduction in the active neutrino population. However, given the lower S$\nu$DM number density compared to that of neutrinos, this modifies $\Delta N_{\rm eff}^\text{\sc cmb}$ by less than 1\%.}
\begin{equation}\label{eq:NeffCMB}
    \Delta N_{\rm eff}^\text{\sc cmb} = 3.046 \left( \frac{\rho_\nu}{\rho_\nu^\text{\sc sc}}-1\right) \ .
\end{equation}
where 
\begin{equation}\label{eq:rhoratio}
\frac{\rho_\nu}{\rho_\nu^\text{\sc sc}} = - \frac{720}{7\pi^4} \left( \frac{T_\nu(z_C)}{T_\nu^\text{\sc sc}(z_C)} \right)^4  {\rm Li_4}(-e^{\mu_\nu(z_C)/T_\nu(z_C)}) \ .
\end{equation}

In short, neutrino self interactions in the early universe via a light mediator $\phi$ imply a tight correlation between the origin of S$\nu$DM and the amount of extra radiation. The post-BBN thermalization of $\phi$ provides a novel mechanism that accounts for the correct relic density of S$\nu$DM. Meanwhile, $\phi$ thermalization distorts the active neutrino phase-space distribution, causing $\Delta N_{\rm eff}>0$ for both CMB and BBN. It is then of great interest to quantify this interplay, especially the prediction for $\Delta N_{\rm eff}^\text{\sc cmb}$, which will serve as a well-motivated target for the upcoming CMB-S4.

\textbf{Results and Discussions --}
We select three benchmark S$\nu$DM masses, $m_4=\left\lbrace2.5,\, 4,\, 7.1\right\rbrace\,$keV, with the correspondingly largest active-sterile mixing parameter, $\sin^22\theta\simeq \left\lbrace5\times 10^{-9},\, 10^{-9}, \, 7\times10^{-11}\right\rbrace$, consistent with current $X$-ray search limits~\cite{Horiuchi:2013noa,Malyshev:2014xqa,Abazajian:2017tcc,Dessert:2018qih}. In Fig.~\ref{fig:LambdaMPhi}, we scan over ($m_\phi$, $\lambda$), focusing on $m_\phi$ between keV and MeV, and derive the black solid curves that can account for the observed DM relic abundance, $\Omega_{\nu_4} h^2=0.1186$. For $m_\phi \gg m_4$, all the curves follow $\lambda \sim m_\phi^{1/2}$, matching the result for $m_\phi\gtrsim1$\,MeV~\cite{deGouvea:2019phk}. As $m_\phi \to m_4$, the curves bend up, where larger coupling $\lambda$ is required, compensating for the phase-space suppression of the $\phi\to\nu\nu_4$ decay. We also present contours of constant $\Delta N_{\rm eff}^\text{\sc bbn}$ and $\Delta N_{\rm eff}^\text{\sc cmb}$ generated by the $\phi$-$\nu$ interaction. Their intersections with the black contours dictate the correlation between $\Omega_{4}$ and $\Delta N_{\rm eff}$.  For real (complex) scalar $\phi$, we find the following predictions 
$0<\Delta N_{\rm eff}^\text{\sc bbn}<0.57\ (1.14)$ and 
$0.12\ (0.19)<\Delta N_{\rm eff}^\text{\sc cmb}<0.9\ (1.95)$.  
The maximal values correspond to strong couplings with $\phi$ fully thermalized by BBN, whereas the minimal values correspond to sufficiently small $\lambda$ and/or $m_\phi$ and negligible pre-BBN population of $\phi$. In the lower-left half of each $\lambda$-$m_\phi$ plane, both $\Delta N_{\rm eff}^\text{\sc bbn}$ and $\Delta N_{\rm eff}^\text{\sc cmb}$ follow the same parametrical dependence as the $\phi$ production rate $\gamma_{\phi \leftrightarrow \nu \nu}$ in  Eq.~(\ref{eq:PhiFreezeIn}) and depend only on the product $(\lambda m_\phi)$. 

Thanks to the late thermalization of $\phi$, the predicted $\Delta N_{\rm eff}^\text{\sc cmb}$ values are always higher than $\Delta N_{\rm eff}^\text{\sc bbn}$. However, we find $\Delta N_{\rm eff}^\text{\sc bbn}$ is not always negligible and it serves as an independent test of non-standard neutrino cosmology. The prior-to-BBN production of $\phi$ particles can increase the expansion rate of the universe, leading to a higher neutron-to-proton ratio when BBN begins, and thus higher helium and deuterium abundances~\cite{Blinov:2019gcj}. 
The state-of-the-art limit, $\Delta N_{\rm eff}^\text{\sc bbn}\lesssim0.5$~\cite{Pitrou:2018cgg}, marginally excludes the case of fully-thermalized real scalar $\phi$ prior to BBN and sets a stronger constraint for complex $\phi$.  Note that the scalar can only mediate neutrino self-interactions, ensuring that weak interactions are unaffected, and hence heat transfer from the plasma to the neutrino sea is only indirectly impacted. The changes produced in the primoridial abundances of helium and deuterium are well within observational uncertainties, and hence the state-of-the-art limit applies~\cite{Grohs:2020xxd}.
Precision calculations of neutrino decoupling taking into account of entropy transfer from the plasma to $\nu$ (and $\phi$) could improve the accuracy of the $\Delta N_{\rm eff}^\text{\sc bbn}$ prediction to $\mathcal{O}(1\%)$ level~\cite{Grohs:2015eua, Grohs:2015tfy}.

\begin{figure*}
 \includegraphics[width=0.8\linewidth]{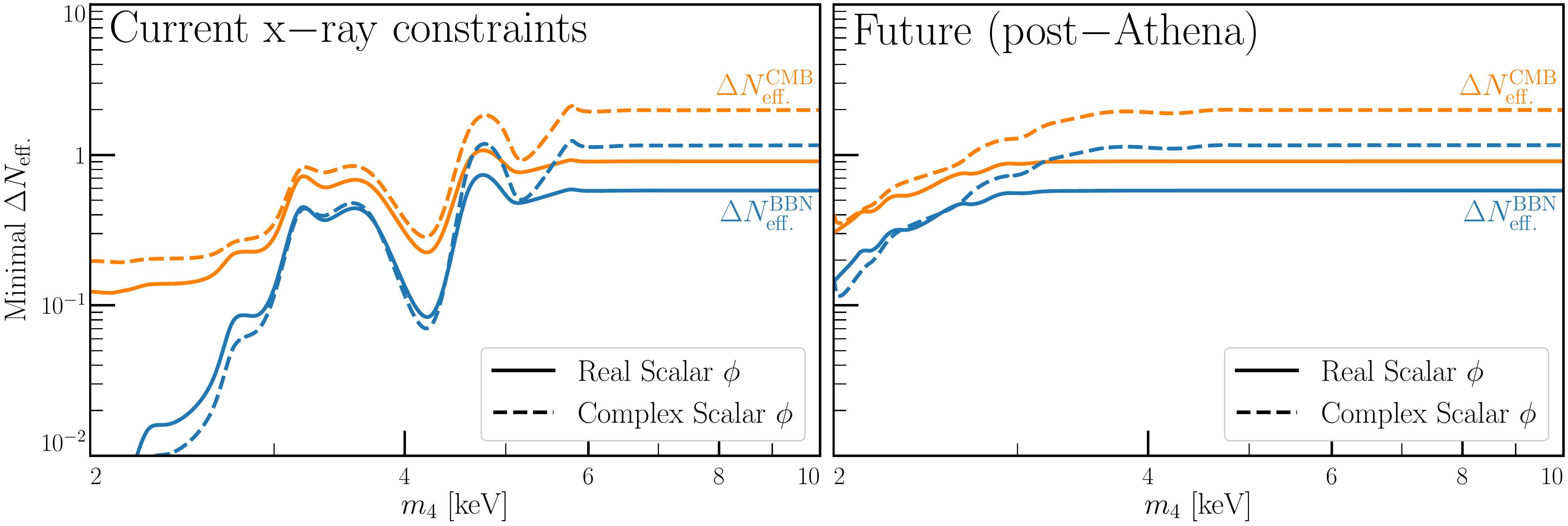} 
    \caption{Minimal value of $\Delta N_{\rm eff}$ at the time of CMB (orange) and BBN (blue) as a function of S$\nu$DM mass $m_{4}$. All contours correspond to the largest allowed active-sterile mixing parameter $\sin^2\left(2\theta\right)$ by the current (left) and upcoming (right) searches of $X$-ray line from $\nu_4\to\nu \gamma$ decay. }
    \label{fig:MinNeff}
\end{figure*}

From Fig.~\ref{fig:LambdaMPhi}, for each S$\nu$DM mass one can derive the lowest predicted values of $\Delta N_{\rm eff}^\text{\sc bbn}$ and $\Delta N_{\rm eff}^\text{\sc cmb}$. Reducing $\theta$ would require larger $\lambda$ to maintain $\Omega_{\nu_4}$, resulting in larger $\Delta N_{\rm eff}^\text{\sc bbn}$ and $\Delta N_{\rm eff}^\text{\sc cmb}$ until they saturate to their maxima. Remarkably, this predicts a target $\Delta N_{\rm eff}$ for precision BBN and CMB measurements.
Moreover, because $\theta$ is more strongly constrained for higher S$\nu$DM mass, an upper bound on $\Delta N_{\rm eff}$ could set an upper bound on $m_4$. This forms a novel interplay with DM indirect detection experiments.
Inspired by this, we present Fig.~\ref{fig:MinNeff}. For each $m_4$, we combine the current $X$-ray constraint and the $\Omega_{\nu_4}$ requirement to derive minimal values of $\Delta N_{\rm eff}^\text{\sc bbn}$ and $\Delta N_{\rm eff}^\text{\sc cmb}$, as shown in the left panel. We refer to supplemental material~B for more details.  We find $\Delta N_{\rm eff}^\text{\sc cmb}>0.12$ for all regions of interest, a favorable target CMB-S4. For $m_4\gtrsim6\,$keV, $\Delta N_{\rm eff}^\text{\sc bbn}$ and $\Delta N_{\rm eff}^\text{\sc cmb}$ have already reached their maxima. Fig.~\ref{fig:MinNeff}(right) shows the impact of projected $\sin^22\theta$ constraints by the Athena $X$-ray observatory~\cite{Neronov:2015kca} on $\Delta N_{\rm eff}$. With smaller $\sin^22\theta$, higher minimal values of $\Delta N_{\rm eff}^\text{\sc bbn}$ and $\Delta N_{\rm eff}^\text{\sc cmb}$, in particular, $\Delta N_{\rm eff}^\text{\sc cmb}>0.3$, are necessary. These predictions will be robustly tested by CMB-S4, with $\Delta N_{\rm eff}^\text{\sc cmb}$ precision at the few-percent level. The two classes of probes are highly complementary and will further restrict or discover S$\nu$DM.

The range of S$\nu$DM mass shown in Fig.~\ref{fig:MinNeff} is $2\,{\rm keV}<m_4<10\,$keV. Lower values of $m_4$ are inconsistent with the phase space density constraint from dwarf galaxies~\cite{Tremaine:1979we, Boyarsky:2008ju, Abazajian:2017tcc}. A stronger lower bound on $m_4$ may be set using Lyman-$\alpha$ forest~\cite{Yeche:2017upn} and Milky Way satellite dwarf galaxy counts~\cite{Nadler:2020prv}. In our scenario, the DM free-streaming length can be substantially shorter than regular warm DM. In numerical calculation, we note that S$\nu$DM is dominantly produced around $z\sim3-4$ where $\phi$ already starts to turn non-relativistic  (see Fig.~2 in supplemental material~A). The initial DM velocity is thus suppressed if $m_\phi \sim m_4$. Interestingly, such a near-degeneracy exactly occurs where $\Delta N_{\rm eff}$ is minimized (see Fig.~\ref{fig:LambdaMPhi}). As a result, small scale constraints are relaxed.

So far we have assumed that $\phi$ couples equally to all active neutrinos. Our conclusions qualitatively hold if this assumption is relaxed to generic flavor non-universal and/or off-diagonal couplings. By the time of $\phi$ thermalization, there is sufficient time for neutrino flavor conversions to occur, which necessarily equilibrates $\phi$ with all the active neutrinos.

Before closing, we comment on other relevant constraints. 
A sub-MeV $\phi$ that couples to the electron neutrino of the form Eq.~(\ref{eq:PhiIntLagrangian}) can affect the two-neutrino double-beta decay rate. The current limit, $\lambda\lesssim10^{-4}$~\cite{Agostini:2015nwa,Blum:2018ljv,Brune:2018sab}, lies above Fig.~\ref{fig:LambdaMPhi}. Another slightly weaker constraint on the $\phi$-$\nu$ coupling comes from the cooling of core-collapsing supernovae~\cite{Raffelt:1996wa, Heurtier:2016otg, Brune:2018sab, Escudero:2020ped}. 

In summary, we explored the impact of active neutrino self-interaction, mediated by an ultralight scalar $\phi$, on the relic density of S$\nu$DM and the amount of extra radiation ($\Delta N_{\rm eff}$) in the early Universe. Strong neutrino self-interactions can facilitate efficient production of DM and generate a net $\Delta N_{\rm eff}$ at the same time. Our study reveals an intimate relationship between these two important quantities and provides a well-motivated target for the future cosmological and astrophysical experiments, including CMB-S4 and Athena. 
These results, together with earlier work, depicts the complete parameter space for S$\nu$DM, with neutrino self-interactions mediated by a scalar whose mass spans over more than six orders of magnitude. They serve as well a motivated target for future experimental tests, from laboratories to the cosmos.
Observations made in this work could extend to other neutrino self-interaction models mediated by vector bosons, such as gauged $U(1)_{L_\mu-L_\tau}$~\cite{Kelly:2020pcy, Shuve:2014doa, Escudero:2019gzq} and other anomaly free gauge theories of baryon and lepton number.


\begin{acknowledgments}
\textbf{Acknowledgements:} We thank Nikita Blinov for discussions and comments on the draft. Fermilab is operated by the Fermi Research Alliance, LLC under contract No. DE-AC02-07CH11359 with the United States Department of Energy. M.S. acknowledges support from the National Science Foundation, Grant PHY-1630782, and to the Heising-Simons Foundation, Grant 2017-228. Y.Z. is supported by the Arthur B. McDonald Canadian Astroparticle Physics Research Institute.
\end{acknowledgments}

\bibliography{references}

\newpage 

\section{Supplemental Material}
\subsection{A. Matching Conditions}\label{appendix1}
In this section, we provide the details on the matching conditions among the three time 
scales defined in the main text, $z_A, z_B, z_C$. This allows us to derive the time $(z)$ 
dependence of the active neutrino temperature $T_\nu(z)$ and chemical potential 
$\mu_\nu(z)$.

At time $z_A$, the active neutrinos have just decoupled from the thermal plasma of other SM particles, thus $T_\nu(z_A) = T_\nu^\text{\sc sc}(z_A) \simeq 1\,$MeV and $\mu_\nu(z_A)=0$. 
The corresponding energy and number densities of active neutrinos are
\begin{equation}
\rho_\nu(z_A) = \frac{7\pi^2}{80} T_\nu(z_A)^4, \quad 
n_\nu(z_A) = \frac{9\zeta(3)}{4\pi^2} T_\nu(z_A)^3 \ .
\end{equation}
Here we count neutrinos and antineutrinos separately, i.e., 
$\rho_{\bar\nu} = \rho_\nu$, $n_{\bar\nu} = n_\nu$.
Meanwhile, there also exist a sub-thermal population of $\phi$ that was frozen in
via its interaction with active neutrinos. The corresponding values, denoted by $\rho_\phi(z_A)$ and $n_\phi(z_A)$,
can be obtained by numerically solving Eq.~(5) of the main text.

\begin{figure}[h]
\centerline{\includegraphics[width=1\linewidth]{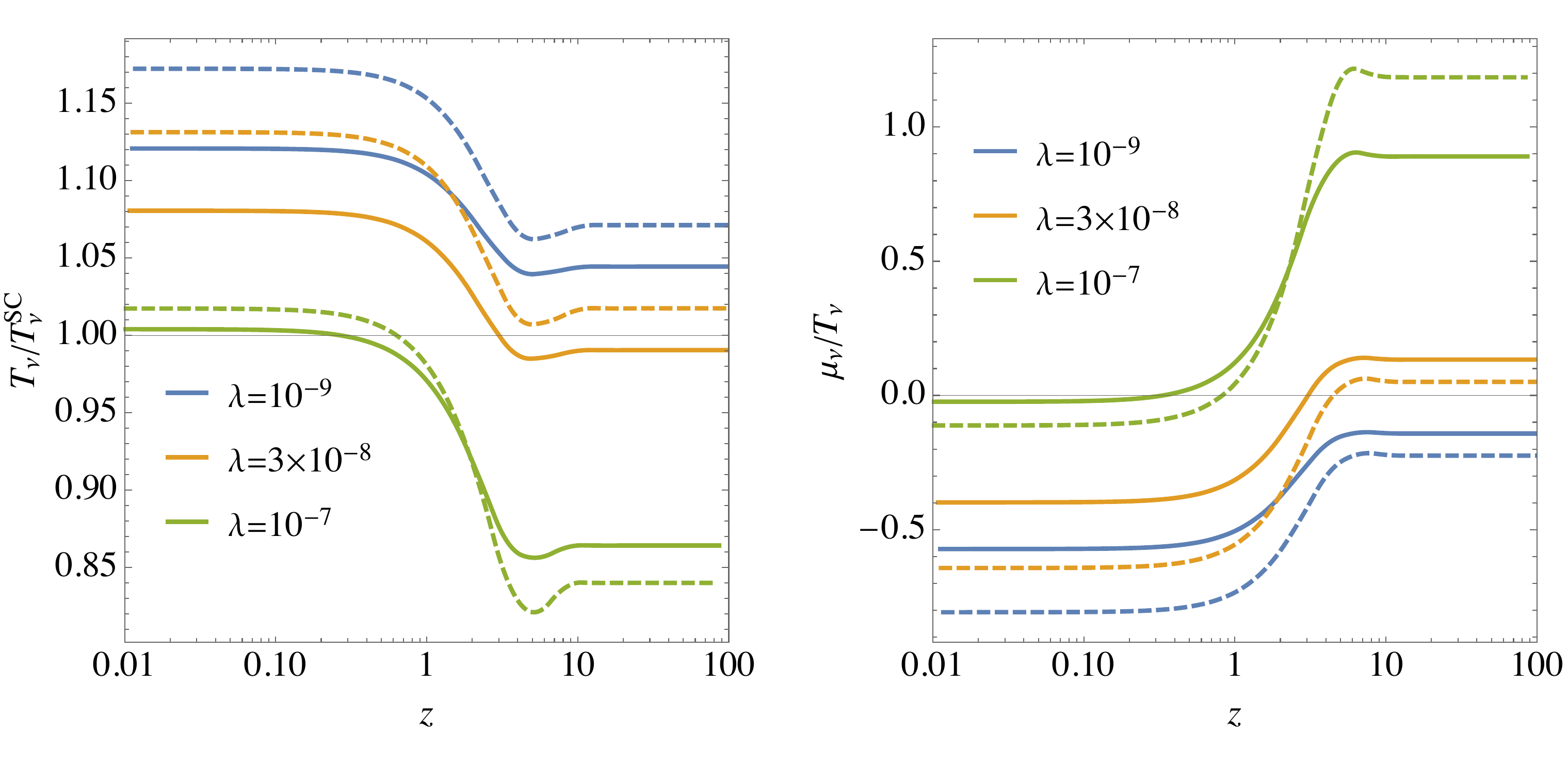}}
    \caption{Evolution of ratios $T_\nu(z)/T_\nu^\text{\sc sc}(z)$ and $\mu_\nu(z)/T_\nu(z)$ as functions of $z$ for three values of $\lambda_\phi$ and holding $m_\phi = 5\,{\rm keV}$ fixed. Solid (dashed) curves correspond to  real (complex) scalar $\phi$ case.}
    \label{fig:TandMu}
\end{figure}

The time $z_B$ is defined when $\phi$ has just entered chemical equilibrium with active neutrinos.
At this moment, the $\phi$-$\nu$ system shares the same temperature $T_\nu(z_B)$, and also develops chemical potential 
where $\mu_\phi(z_B) = 2 \mu_\nu(z_B)$. The corresponding energy, number, and entropy densities are
\begin{equation}\label{distributionB}
\begin{split}
\rho_\nu(z_B) &= - \frac{9T_\nu(z_B)^4}{\pi^2} {\rm Li}_4(-e^{\mu_\nu(z_B)/T_\mu(z_B)}) \ , \\
n_\nu(z_B) &= - \frac{3T_\nu(z_B)^3}{\pi^2} {\rm Li}_3(-e^{\mu_\nu(z_B)/T_\mu(z_B)}) \ , \\
s_\nu(z_B) &= -\frac{12T_\nu(z_B)^3}{\pi^2} {\rm Li}_4(-e^{\mu_\nu(z_B)/T_\mu(z_B)}) \\
&\hspace{0.4cm} + \frac{3\mu_\nu(z_B) T_\nu(z_B)^2}{\pi^2}  {\rm Li}_3(-e^{\mu_\nu(z_B)/T_\mu(z_B)}) \ , \\
\rho_\phi(z_B) &= \frac{3T_\nu(z_B)^4}{\pi^2} {\rm Li}_4(e^{2\mu_\nu(z_B)/T_\mu(z_B)}) \ , \\
n_\phi(z_B) &= \frac{T_\nu(z_B)^3}{\pi^2} {\rm Li}_3(e^{2\mu_\nu(z_B)/T_\mu(z_B)}) \ , \\
s_\phi(z_B) &= \frac{4T_\nu(z_B)^3}{\pi^2} {\rm Li}_4(e^{2\mu_\nu(z_B)/T_\mu(z_B)}) \\
&\hspace{0.4cm} - \frac{2\mu_\nu(z_B) T_\nu(z_B)^2}{\pi^2}  {\rm Li}_3(e^{2\mu_\nu(z_B)/T_\mu(z_B)}) \ , \\
\end{split}
\end{equation}
where we treat $\phi$ as an ultra-relativistic species ($m_\phi\ll T_\nu$), which is a good approximation 
throughout the parameter space we explore in this work.
Again, in the case where $\phi$ is a complex scalar, the densities of the $\phi^*$ degree of freedom are counted separately. In the absence of CP violation,
$\rho_{\phi^*} = \rho_\phi$, $n_{\phi^*} = n_\phi$, $s_{\phi^*}= s_\phi$.

The matching conditions between $z_A$ and $z_B$ are dictated by energy and lepton number conservation~\cite{Chacko:2003dt}.
If $\phi$ is a real scalar, it comes to chemical equilibrium with both $\nu$ and $\bar\nu$, thus
\begin{equation}\label{A2Breal}
\begin{split}
\frac{\rho_\nu(z_A) + \rho_{\bar\nu}(z_A) + \rho_\phi(z_A)}{\rho_\nu(z_B) + \rho_{\bar\nu}(z_B) + \rho_\phi(z_B)} &= \frac{z_B^4}{z_A^4} \ , \\
\frac{n_\nu(z_A) +n_{\bar\nu}(z_A) + 2n_\phi(z_A)}{n_\nu(z_B) + n_{\bar\nu}(z_B) + 2n_\phi(z_B)} &= \frac{z_B^3}{z_A^3} \ . \\
\end{split}
\end{equation}
Alternatively, if $\phi$ is a complex scalar, it only comes into equilibrium with neutrinos ($\phi^*$ equilibrates with $\bar\nu$), thus
\begin{equation}\label{A2Bcomplex}
\begin{split}
\frac{\rho_\nu(z_A) + \rho_\phi(z_A)}{\rho_\nu(z_B)  + \rho_\phi(z_B)} &= \frac{z_B^4}{z_A^4} \ , \\
\frac{n_\nu(z_A) + 2n_\phi(z_A)}{n_\nu(z_B) + 2n_\phi(z_B)} &= \frac{z_B^3}{z_A^3} \ . \\
\end{split}
\end{equation}
By definition $z_A/z_B = T_\nu^\text{\sc sc}(z_B)/T_\nu^\text{\sc sc}(z_A)$, and noting that $T_\nu(z_A)=T_\nu^\text{\sc sc}(z_A)$, 
the above set of equations (\ref{A2Breal}) or (\ref{A2Bcomplex}) allows us to derive the ratios
\begin{equation}\label{step1}
\frac{T_\nu(z_B)}{T_\nu^\text{\sc sc}(z_B)}, \quad \frac{\mu_\nu(z_B)}{T_\nu(z_B)} \ .
\end{equation}
In particular, in the case of real $\phi$ and with a negligible population of $\phi$ at time $z_A$ (corresponds to a tiny $\lambda$), we can derive
\begin{equation}\label{example1}
T_\nu(z_B)/T_\nu^\text{\sc sc}(z_B)\simeq1.12, \quad \mu_\nu(z_B)/T_\nu(z_B)\simeq-0.57 \ ,
\end{equation}
consistent with the findings of Ref.~\cite{Escudero:2020dfa}.

\bigskip
As the next step, we consider the evolution of $T_\nu$ and $\mu_\nu$ between $z_B$ and $z_C$, where $z_C$ is defined as a temperature much lower than $m_\phi$. During this epoch, the $\nu$-$\phi$ system remains in chemical equilibrium. As a result, we can describe their thermal distributions at any time $z$ where $z_B<z<z_C$. Because neutrino masses are still negligible, their energy, number and entropy densities are described by the same functions as those in Eq.~(\ref{distributionB}), by simply replacing $z_B\to z$. On the other hand, the mass of $\phi$ becomes important and the corresponding distribution functions are evaluated using
\begin{equation}
\begin{split}
\rho_\phi(z, m_\phi) &= \frac{T_\nu(z)^4}{2\pi^2} \int_{\frac{m_\phi}{T_\nu(z)}}^\infty \frac{\left[x^2 - (m_\phi/T_\nu(z))^2 \right]^{1/2} x^2 dx}{e^{x-\mu_\phi(z)/T_\nu(z)} -1} \ , \\
n_\phi(z, m_\phi) &= \frac{T_\nu(z)^3}{2\pi^2} \int_{\frac{m_\phi}{T_\nu(z)}}^\infty \frac{\left[x^2 - (m_\phi/T_\nu(z))^2 \right]^{1/2} x dx}{e^{x-\mu_\phi(z)/T_\nu(z)} -1} \ , \\
p_\phi(z, m_\phi) &= \frac{T_\nu(z)^4}{2\pi^2} \int_{\frac{m_\phi}{T_\nu(z)}}^\infty \frac{\left[x^2 - (m_\phi/T_\nu(z))^2 \right]^{3/2} dx}{e^{x-\mu_\phi(z)/T_\nu(z)} -1} \ , \\
s_\phi(z, m_\phi) &= \frac{\rho_\phi(z, m_\phi) + p_\phi(z, m_\phi)}{T_\nu(z)} - \mu_\phi(z) n_\phi(z, m_\phi) \ ,
\end{split}
\end{equation}
where $\mu_\phi = 2\mu_\nu$ still holds.

Applying the matching conditions between $z_B$ and $z$ ($z_B<z<z_C$), which are entropy and lepton number conservation
\begin{equation}\label{A2BrealzB}
\begin{split}
\frac{s_\nu(z_B) + s_{\bar\nu}(z_B) + s_\phi(z_B)}{s_\nu(z) + s_{\bar\nu}(z) + s_\phi(z, m_\phi)} &= \frac{z^3}{z_B^3} \ , \\
\frac{n_\nu(z_B) + n_{\bar\nu}(z_B) + 2n_\phi(z_B)}{n_\nu(z) +n_{\bar\nu}(z) + 2n_\phi(z, m_\phi)} &= \frac{z^3}{z_B^3} \ , \\
\end{split}
\end{equation}
for the case of a real $\phi$, or
\begin{equation}\label{A2BcomplexzB}
\begin{split}
\frac{s_\nu(z_B) + s_\phi(z_B)}{s_\nu(z) + s_\phi(z, m_\phi)} &= \frac{z^3}{z_B^3} \ , \\
\frac{n_\nu(z_B) + 2n_\phi(z_B)}{n_\nu(z) + 2n_\phi(z, m_\phi)} &= \frac{z^3}{z_B^3} \ , \\
\end{split}
\end{equation}
for the complex $\phi$, and using the relation $z_B/z = T_\nu^\text{\sc sc}(z)/T_\nu^\text{\sc sc}(z_B)$, we will able to derive the ratio
$\mu_\nu(z)/T_\nu(z)$, and $r(z)$ which is defined as
\begin{equation}
\frac{T_\nu(z)}{T_\nu^\text{\sc sc}(z)} = r(z) \frac{T_\nu(z_B)}{T_\nu^\text{\sc sc}(z_B)} \ .
\end{equation}
Combining this result with the findings in Eq.~(\ref{step1}), we can derive the ratio of neutrino temperatures in this model to that in standard cosmology, $T_\nu(z)/T_\nu^\text{\sc sc}(z)$.

\begin{figure}[!t]
\includegraphics[width=0.8\linewidth]{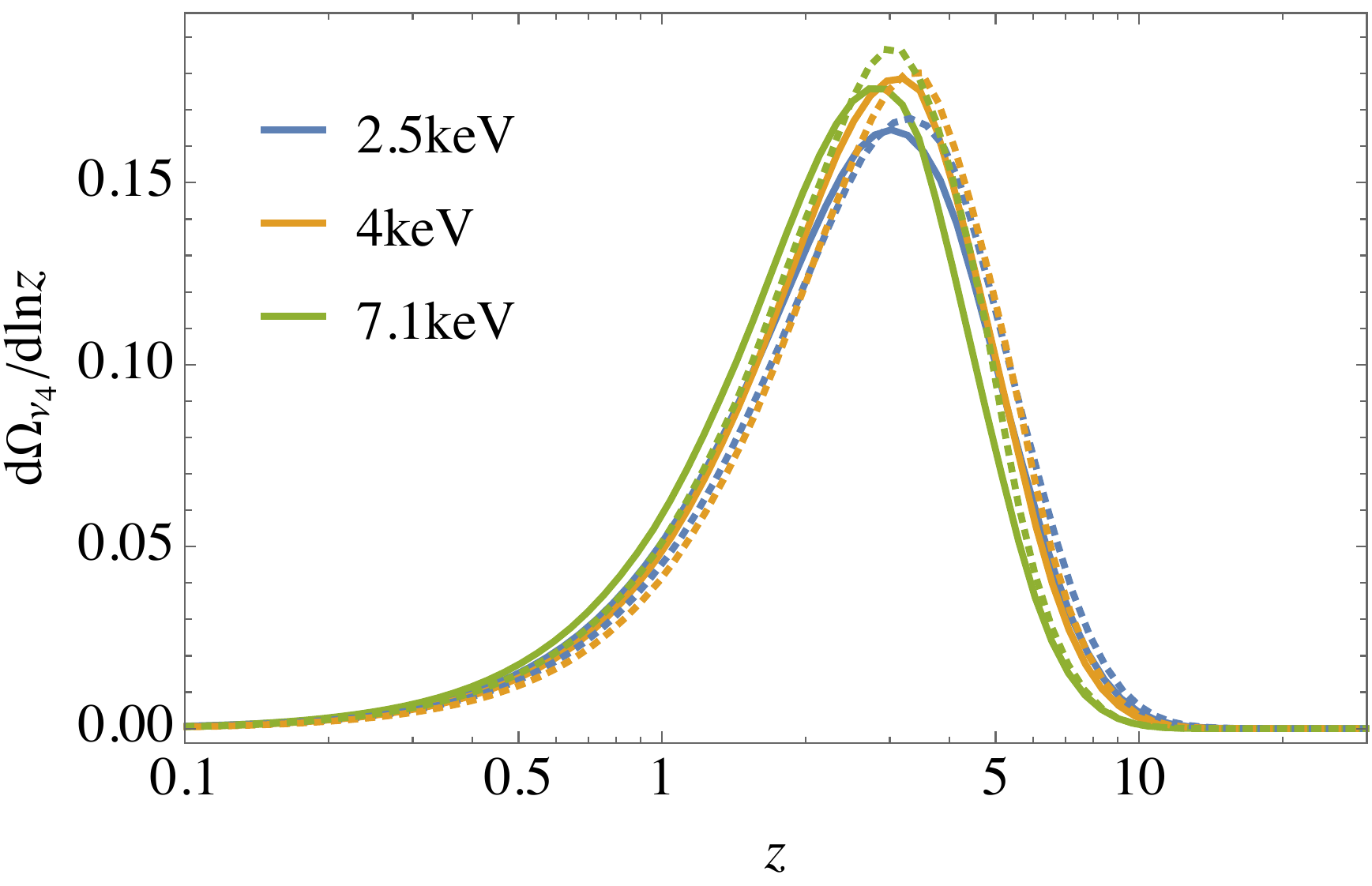}
    \caption{Time dependence of S$\nu$DM, for three values of $m_4$ as labelled. The other parameters are chosen for producing the observed DM relic density. The solid (dashed) curves correspond to real (complex) scalar $\phi$ case.}
    \label{fig:Production}
\end{figure}

For example, in the limit where $z=z_C\gg1$ such that $\rho_\phi, n_\phi, s_\phi\to0$ are all Boltzmann suppressed, for real scalar $\phi$ and negligible population of $\phi$ at $z_A$, we obtain
\begin{equation}\label{example2}
T_\nu(z_C)/T_\nu^\text{\sc sc}(z_C)\simeq1.04, \quad \mu_\nu(z_C)/T_\nu(z_C)\simeq-0.14 \ ,
\end{equation}
also consistent with Ref.~\cite{Escudero:2020dfa}.

In Fig.~\ref{fig:TandMu}, we show the evolution of $T_\nu$ and $\mu_\nu$ as functions of $z$ for three sets of model parameters.
For very small $\lambda$, the asymptotic values are consistent with the results in Eq.~(\ref{example1}) and (\ref{example2}).
For sufficiently large $\lambda$, the freeze in population of $\phi$ (via Eq.~(5) of the main text) before BBN 
is non-negligible which affects the matching results from $z_A$ to $z_C$.
The trend is also consistent with the limit where $\phi$ already thermalizes before BBN. In that case, we would have $z_A=z_B$ and $\mu_\nu(z_B)=0$.

\begin{figure}[!t]
\centerline{\includegraphics[width=0.9\linewidth]{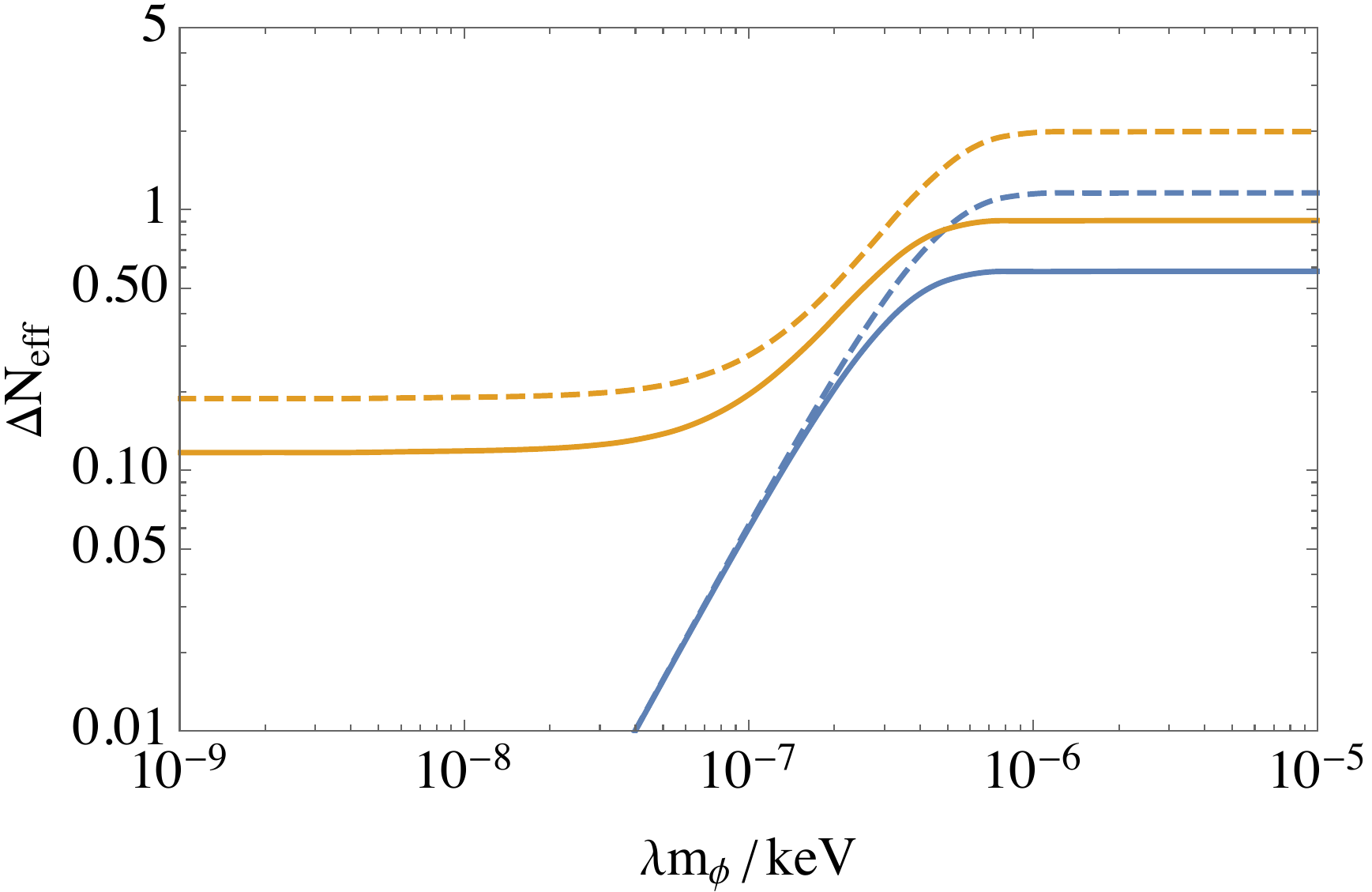}}
    \caption{Parametrical dependence of $\Delta N_{\rm eff}^\text{\sc cmb}$ (orange) and $\Delta N_{\rm eff}^\text{\sc bbn}$ (blue) on the product $\lambda m_\phi$, for real (solid) and complex (dashed) scalar cases. This result is valid as long as $m_\phi\ll 1\,$MeV. }
    \label{fig:CMBBBN}
\end{figure}

With the above results, we calculate the S$\nu$DM relic density using Eq.~(7) in the main text.
In Fig.~\ref{fig:Production}, we plot the time dependence of $d\Omega_{\nu_4}/d\ln z$ for a few sets of model parameters. 
In all cases, we find that S$\nu$DM is dominantly produced around $z\simeq 3-4$ where $\phi$ already starts to become non-relativistic. If $\phi$ and $\nu_4$ is nearly degenerate ($m_\phi\gtrsim m_4$), S$\nu$DM will be produced with a velocity $v\ll c$.

Furthermore, the above results also allow us to calculate $\Delta N_{\rm eff}$ at CMB time. After $\phi$ decays away at $z_C\gg 1$, the ratios 
$T_\nu(z_C)/T_\nu^\text{\sc sc}(z_C)$ and $\mu_\nu(z_C)/T_\nu(z_C)$ approach to constant values throughout the later evolution of the Universe, as shown in Fig.~\ref{fig:TandMu}. We plug these functions into Eq.~(11) of the main text.
In the case of negligible $\phi$ population by the time of BBN, the value of $\Delta N_{\rm eff}^\text{\sc cmb}$ is
0.12 and 0.19 for real and complex $\phi$ case, respectively. The result for the real scalar case is consistent with the findings in Refs.~\cite{Chacko:2003dt}.

We observe that $\Delta N_{\rm eff}^\text{\sc bbn}$ and $\Delta N_{\rm eff}^\text{\sc cmb}$ depend on model parameter space $\lambda$ and $m_\phi$ only through
the combination $\lambda m_{\phi}$ as long as the product is small (corresponding to the low-left half of Fig.~2 of the main text). This dependence and the parametrical correlation are shown in Fig.~\ref{fig:CMBBBN} for real and complex scalars. We see that for sufficiently small $\lambda m_\phi$, there is no impact on $\Delta N_{\rm eff}^\text{\sc bbn}$ (blue curves) -- no abundance of $\phi$ has developed by the time of BBN. On the contrary, even for $\lambda m_\phi \approx 10^{-9}$ keV, $\Delta N_{\rm eff}^\text{\sc cmb}$ has a lowest value, 0.12 or 0.19, depending on whether $\phi$ is real or complex due to distortions of the active neutrino distributions. For both BBN and CMB, large enough $\lambda m_\phi$ results in a complete saturation of $\Delta N_{\rm eff}$ and these curves level off to their maxima.
This approximation is only valid for $m_\phi \ll 1$ MeV, where $\phi$ cannot decay away before BBN.
\subsection{B. Minimal $\Delta N_{\rm eff}$}\label{appendix2}
\begin{figure}[!t]
\includegraphics[width=0.9\linewidth]{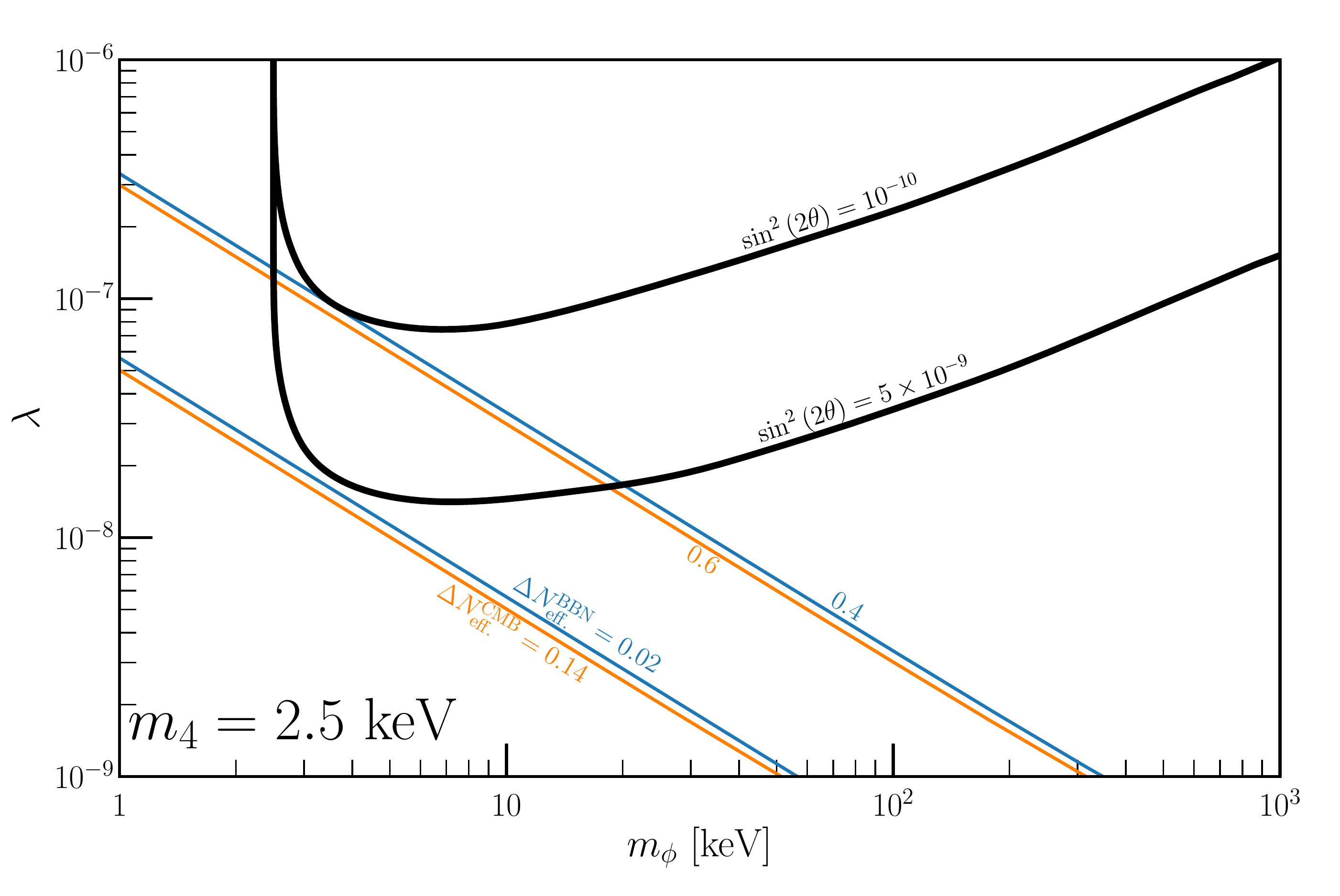}
    \caption{Similar to Fig.~2 of the main text but with $m_4=2.5\,$keV, $\sin\theta=5\times10^{-9}$ and $10^{-10}$ for DM relic density calculation (black curves). The corresponding lowest values of $\Delta N_{\rm eff}^\text{\sc cmb}$ and $\Delta N_{\rm eff}^\text{\sc bbn}$ are shown by the orange and blue curves, respectively. Here we consider the real scalar $\phi$ case.}
    \label{fig:LambdaMPhiAthena}
\end{figure}
Here we provide some further context for the procedure used in generating Fig.~3 of the main text. For a given $m_4$, we determine the largest allowed $\sin^2\left(2\theta\right)$ assuming current limits from $X$-ray searches~\cite{Horiuchi:2013noa,Malyshev:2014xqa,Abazajian:2017tcc,Dessert:2018qih}, as well as the projected upper limit by Athena~\cite{Neronov:2015kca}. For a concrete choice, let us consider $m_4 = 2.5$ keV. The current limit on this mass is $\sin^2\left(2\theta\right) < 5 \times 10^{-9}$, whereas Athena will have sensitivity at the level of $\sin^2\left(2\theta\right) = 10^{-10}$. We determine, for those two values of $\sin^2\left(2\theta\right)$, the preferred region of $\lambda$ vs. $m_\phi$ parameter space for which $\nu_4$ can account for all of the DM observed today -- this is shown as the two solid black lines in Fig.~\ref{fig:LambdaMPhiAthena}. Each of those curves, near the minimum in $\lambda$ realized, attains some minimum value of both $\Delta N_{\rm eff}^\text{\sc cmb}$ and $\Delta N_{\rm eff}^\text{\sc bbn}$. For the case we have considered here, $m_4 = 2.5$ keV, we find that with current limits on $\sin^2\left(2\theta\right)$, $\Delta N_{\rm eff}^\text{\sc bbn}$ ($\Delta N_{\rm eff}^\text{\sc cmb}$) must be at least 0.02 (0.14), and with future constraints (assuming Athena does not observe a signal compatible with $m_4 = 2.5$ keV) will require $\Delta N_{\rm eff}^\text{\sc bbn} > 0.4$ ($\Delta N_{\rm eff}^\text{\sc cmb} > 0.6$). In Fig.~\ref{fig:LambdaMPhiAthena} we have assumed $\phi$ to be a real scalar, but the procedure is the same for complex $\phi$. These minimum values are what is shown for that choice of $m_4$ in the left (current constraint on $\sin^2\left(2\theta\right)$) and right (post-Athena constraint) panels of Fig.~\ref{fig:MinNeff}.
\end{document}